\documentclass[aps,pra,twocolumn,superscriptaddress,showpacs]{revtex4}

\usepackage{graphicx}

\usepackage{array}

\def\be{\begin{equation}}
\def\ee{\end{equation}}
\def\bea{\begin{eqnarray}}
\def\eea{\end{eqnarray}}
\def\ba{\begin{array}}
\def\ea{\end{array}}
\def\bdm{\begin{displaymath}}
\def\edm{\end{displaymath}}

\begin{document}



\title{Theory of SU(N) Fermi liquid
\\
}


\author{S.-K. Yip}
\affiliation{Institute of Physics, Academia Sinica, Taipei 115, Taiwan}
\affiliation{Institute of Atomic and Molecular Sciences, Academia Sinica, Taipei 106, Taiwan}
\author{Bor-Luen Huang}
\affiliation{Institute of Physics, Academia Sinica, Taipei 115, Taiwan}
\author{Jung-Shen Kao}
\affiliation{Institute of Physics, Academia Sinica, Taipei 115, Taiwan}
\affiliation{Department of Physics, National Taiwan University, Taipei 106, Taiwan}



\date{\today}

\begin{abstract}
We generalized the Fermi liquid theory to
$N$ component systems with SU(N) symmetry.
We emphasize the important role of fluctuations
when $N$ is large.  These fluctuations dramatically modifies
the properties for repulsive Fermi gases, in particular the
spin susceptibility.

\end{abstract}

\pacs{03.75.Ss, 67.85.Lm, 67.85.-d}

\maketitle



\section{Introduction}

Recent advances in the field of cold atoms provide us with many new opportunities
to study quantum multiparticle physics \cite{review1,review2}, in systems or with methods often
not available in traditional condensed matter settings.
In this paper, we consider one such example, namely
fermionic systems with SU(N) symmetry, with $N > 2$.  This is naturally
available in cold atomic gases of alkaline atoms \cite{Yb,Sr,Yb-m,Sr-m1,Sr-m2,Sr-m3,Yb-Mott} without much fine tuning
\cite{CHU,Hermele09,Gorshkov10}.  With two electrons in the outermost shell and in the total
spin and orbital angular momentum zero state, the hyperfine spin $f$ of the atom is due
entirely to the nuclear spin, which however basically has no influence on the
particle-particle scattering length which parameterizes the effective interaction
in the dilute, low energy limit.  The effective interaction is thus automatically
SU(N) symmetric, with $N$ the number of species trapped.
If all sublevels are trapped, then $N=2f +1$.
$N$ can also be any counting number less than $(2f +1)$, if only a subset of
them is trapped \cite{Sr-m1,Sr-m2,Sr-m3}.
$N$ can thus be $> 2$ if $f > 1/2$.
Recently, a lot of experimental progress has been made on these systems.
In particular both the Fermi gases $^{173}$Yb with $f = 5/2$ and
$^{87}$Sr with $f =9/2$ have been cooled to quantum degeneracy
\cite{Yb,Sr}.
The Mott state of such a multi-component system has already been reached \cite{Yb-Mott}.
Many theorists have predicted that this Mott state should exhibit
 exotic ordering at low temperatures\cite{Hermele09,Gorshkov10,SUNm1,SUNm2,SUNm3,SUNm4,SUNm5}.
 Also available are degenerate Bose-Fermi mixtures \cite{Sr-m1,Sr-m2}
and Fermi-Fermi mixtures \cite{Yb-m}.  This latter system has
SU(2) $\times$ SU(6) symmetry and can in principle hosts a very
interesting superfluid\cite{Yip11}, though this regime is
yet to be achieved.

In this paper, we study one aspect of the SU(N) fermionic systems which
should be readily accessible experimentally, namely their Fermi liquid properties.
In its original form \cite{Landau1,Landau2,PN}, the Landau Fermi liquid
theory was intended for a two-component system with SU(2) symmetry.
We generalize the theory to SU(N).
We evaluate the Fermi liquid parameters,
in particular in the dilute limit.   We obtain
the compressibility
and the generalized magnetic susceptibilities of this gas.  These
quantities can be readily measured by monitoring the density profile
and relative number fluctuations.  We illuminate in particular
how these quantities depend on the number of components.
We show that an increase in this number leads to a dramatic
suppression of the spin susceptibility of repulsive gases.  This is
a consequence of an enhanced effective repulsion between identical
particles, generated by induced density fluctuations of the other species.

\section{Formulation}

The Hamiltonian of our system is given by

\bea
H &=& \sum_{\vec k \gamma} \left( \frac{k^2}{2M} - \mu \right)
 a_{\vec k \gamma}^{\dagger} a_{\vec k \gamma}  + \nonumber \\
 & & \frac{g}{2} \sum_{\vec k_1 \vec k_2 \vec k_3 \vec k_4, \gamma_1 \ne \gamma_2}
  a_{\vec k_1 \gamma_1}^{\dagger}  a_{\vec k_2 \gamma_2}^{\dagger} a_{\vec k_3 \gamma_2} a_{\vec k_4 \gamma_1}
  \delta_{\vec k_1 + \vec k_2, \vec k_3 + \vec k_4}
  \label{H}
  \eea
where the first term is the kinetic energy and the second, the interaction.
$ a_{\vec k \gamma}^{\dagger}$ and $ a_{\vec k \gamma} $ are the creation and
annihilation operators for a Fermion of the $\gamma$ species with momentum $\vec k$,
$M$ is the atomic mass, and $\mu$ is the chemical potential, assumed identical
for all the species at the moment.  $g$ characterizes the short range two-body interaction,
which will  be eliminated in favor  of the s-wave scattering length between the particles.
In the interaction term, it is sufficient to include $\gamma_1 \ne \gamma_2$ due
to Fermionic antisymmetry.  $g$ is independent of the species involved, so
we do  not have spin conversion processes in the sense that the outgoing particles
must have the same spin indices $\gamma_1$, $\gamma_2$ as the incoming particles.
Eq (1) can be obtained from the more general Hamiltonian in \cite{YipHo} if
we set all the scattering lengths there to be equal.
Hereafter, to avoid possible confusion with the particle number, the number
of species will be denoted as $N_c$.
Eq (1) has SU($N_c$) symmetry
since it is invariant under any unitary transformation
($a_\alpha \to \sum_{\alpha'} U_{\alpha \alpha'} a_{\alpha'}$ with
$U_{\alpha \alpha'}$ unitary) among the $N_c$ species involved.

In Fermi liquid theory, the low energy Fermionic excitations are
described by quasiparticle excitations $\delta n_{\delta \gamma} (\vec k)$ at $\vec k$.
$\delta n_{\delta \gamma} (\vec k)$ is an $N_c \times N_c$  matrix in the species indices.
The change in quasiparticle energy at $\vec k$ is $\delta \epsilon_{\alpha \beta} (\vec k)$,
also an $N_c \times N_c$ matrix, related to $\delta n_{\delta \gamma}$ by

\be
\delta \epsilon_{\alpha \beta} (\vec k) = \sum_{\vec k' \gamma \delta}
f_{\alpha \gamma, \beta \delta} (\vec k, \vec k') \delta n_{\delta \gamma} (\vec k')
\label{def-f}
\ee
where it is understood that all $\vec k$, $\vec k'$ are
near the Fermi surface. Our notations are generalization of those in \cite{LL} from two to $N_c$ species.
Slightly different notations were used in \cite{CHU}.  Ref \cite{CHU} parameterized
$f_{\alpha \gamma, \beta \delta}$ as

\be
f_{\alpha \gamma, \beta \delta} (\vec k, \vec k') =
f_s (\vec k, \vec k') \delta_{\alpha \beta} \delta_{\gamma \delta}
 + 4 f_m (\vec k, \vec k') \sum_a T^a_{\alpha \beta} T^a_{\gamma \delta}
 \label{para-f1}
\ee
where the matrices ${\bf T^a}$, $a = 1, \ldots, N_c^2 - 1$ are the (traceless)
generators of $SU(N_c)$ with the normalization ${\rm Tr} [{\bf T^a T^b}] = \frac{1}{2} \delta^{a b}$,
and ${\rm Tr}$ denotes the trace.  Here, we have defined $f_m$ which differs from \cite{CHU}
by a factor of $2$ so that eq (\ref{para-f1}) will reduce to the form
of \cite{LL} if $N_c =2 $ where ${\bf T^a}$ will become ${\bf \sigma^a}/2$ with
${\bf \sigma^a}$ being the Pauli matrices.
Eq (\ref{para-f1}) implies that, if we write
\be
\delta n_{\delta \gamma} (\vec k) = \frac{\delta n (\vec k)}{N_c} \delta_{\delta \gamma}
  + \frac{1}{2} \delta m_{\delta \gamma}
  \label{def-mm}
  \ee
with
\be
\delta m_{\delta \gamma} =  \sum_a m^a T^a_{\delta \gamma}
\label{def-ma}
\ee
 and
\be
\delta \epsilon_{\alpha \beta} (\vec k) = \delta \epsilon_s (\vec k) \delta_{\alpha \beta}
  - \delta h_{\alpha \beta}
  \label{def-hm}
\ee
with
\be
\delta h_{\alpha \beta} = \sum_a \delta h^a T^a_{\alpha \beta} \ ,
\label{def-ha}
\ee
that is, we separate out, for both $\delta n_{\alpha \beta}$ and $\delta \epsilon_{\alpha \beta}$
the parts that are proportional to the identity matrix and those which are linear combinations
involving ${\bf T^a}$'s , we get
\be
\delta \epsilon_s (\vec k) = \sum_{\vec k'} f_s (\vec k, \vec k') \delta n(\vec k')
\label{esn}
\ee
and
\be
\delta h^a (\vec k) = - \sum_{\vec k'} f_m (\vec k, \vec k') m^a (\vec k')
\label{hm}
\ee
The above form is suggested by the fact
that $\delta n (\vec k) = \sum_{\gamma} \delta n_{\gamma \gamma} (\vec k)$ is
 the total density change at $\vec k$, and $\delta m^a (\vec k)$
is a generalized magnetization.  Under a unitary transformation among
the components, $\delta n (\vec k)$ is invariant whereas $\delta m^a$ would transform
among each other. Similarly, $\delta \epsilon_s (\vec k)$ is the part
of $\delta \epsilon_{\alpha \beta}$ that is invariant under SU($N_c$), whereas
$\delta h^a (\vec k)$ is a kind of generalized magnetic field.

For our later purposes, we shall instead use
\bea
f_{\alpha \gamma, \beta \delta} (\vec k, \vec k')
 &=& \left(f_s (\vec k, \vec k') - \frac{2}{N_c} f_m (\vec k, \vec k') \right)
\delta_{\alpha \beta} \delta_{\gamma \delta}  \nonumber \\
 & & + 2 f_m (\vec k, \vec k') \delta_{\alpha \delta} \delta_{\beta \gamma}
\label{para-f2}
\eea
That this parametrization is possible is suggested by the fact
that
$\delta_{\alpha \beta} \delta_{\gamma \delta}$ and $\delta_{\alpha \delta} \delta_{\beta \gamma}$
are two linearly independent invariants under the SU($N_c$) transformations,
and eq (\ref{para-f1}) also just contains two such invariants.
Indeed, it is simple to verify that, when eq (\ref{def-mm}-\ref{def-ha}) are
substituted in eq (\ref{para-f2}), we obtain the same $\delta \epsilon_s (\vec k)$ and
$\delta h^a(\vec k)$ as in eq (\ref{esn}) and (\ref{hm}).
Note that we simply have
$\delta h_{\alpha \beta} (\vec k) =  - \sum_{\vec k'} f_m (\vec k, \vec k') \delta m_{\alpha \beta} (\vec k')$.

Eq (\ref{para-f2}) is particularly easy to understand if we consider the special
case where all distribution functions are diagonal,  that is,
$\delta n_{\delta \gamma} (\vec k) = \delta n_{\gamma} (\vec k) \delta_{\delta \gamma}$
($\gamma$ not summed).  Then $\delta \epsilon_{\alpha \beta} (\vec k) $ is also diagonal and
can be written as $\delta \epsilon_{\alpha} \delta_{\alpha \beta}$.
 By SU($N_c$) symmetry,
we expect
\be
\delta \epsilon_{\alpha} (\vec k) = \sum_{\vec k'} f_{\alpha \alpha} (\vec k, \vec k')
 \delta n_{\alpha} (\vec k') +
 \sum_{\vec k' \beta \ne \alpha} f_{\alpha \beta \ne \alpha} (\vec k, \vec k')
 \delta n_{\beta} (\vec k')
 \label{d}
 \ee
 with $f_{\alpha \alpha}$ independent of $\alpha$ and
 $f_{\alpha \beta \ne \alpha}$ independent of the species involved
 (so long as they are distinct).
 $f_{\alpha \alpha}$ and $f_{\alpha \beta \ne \alpha}$ play
 the role of effective interaction between identical and distinguishable
 species, respectively.
 Indeed, eq (\ref{para-f2}) and eq (\ref{d}) are equivalent if
 \bea
 f_{\alpha \alpha} (\vec k, \vec k') &=& f_s (\vec k, \vec k') +
   2 \left( 1 - \frac{1}{N_c} \right) f_m (\vec k, \vec k')
   \label{faa}
   \\
f_{\alpha \beta \ne \alpha} (\vec k, \vec k')
   &=& f_s (\vec k, \vec k') - \frac{2}{N_c} f_m (\vec k, \vec k')
   \label{fab}
   \eea
  We shall make use of these expressions below.

  With the Landau liquid formulation, it is straight-forward to
  evaluate the response of the system to external perturbations.
  Consider uniform, $\vec k$ independent external potentials parameterized
  in the form
$  \delta \epsilon^{\rm ext}_{\alpha \beta}
   = \delta \epsilon_s^{\rm ext} \delta_{\alpha \beta} -\delta h^{\rm ext}_{\alpha \beta} $
   with
$   \delta h^{\rm ext}_{\alpha \beta} = \sum_a h^{a, {\rm ext}} T^a_{\alpha \beta}$
  in analogy with eq (\ref{def-hm}) and (\ref{def-ha}). One can
  easily show that (see Supplemental Material \cite{Supp}) the
   density change $\delta n = \sum_{\vec k} \delta n(\vec k)
   \equiv \left( \frac{dn}{d \epsilon_s^{\rm ext}} \right) \delta \epsilon_s^{\rm ext}$
   and is thus linear in $\delta \epsilon_s^{\rm ext}$ and independent of $\delta h^{a {\rm ext}}$,
   and similarly
  $ m^a = \chi \delta h^{a, {\rm ext}}$ independent of $\delta \epsilon_s^{\rm ext}$.
  Since a uniform increase in energy for all the species is equivalent to
  a lowering of the chemical potential $\mu$,  we shall write
  $ \frac{dn}{d \epsilon_s^{\rm ext}} = - \frac{d n}{d \mu}$.
  We obtain
  \be
  \frac{d n}{d \mu} = \frac{N_c N(0)}{ 1 + N_c N(0) f_{0,s}}
  \label{comp}
  \ee
  and
  \be
  \chi = \frac{2 N(0)}{ 1 + 2 N(0) f_{0,m}}
  \label{chi}
  \ee
  Here $f_{0,s}$ and $f_{0,m}$ are the angular averages of
  $f_s(\vec k, \vec k')$ and $f_m (\vec k, \vec k')$ over the Fermi surface,
  and $N(0)$ is the density of states at the Fermi level for a single species,
  given by $\frac{M^* k_F}{2 \pi^2}$ where $M^*$ is the effective mass of
  the quasiparticles, and $k_F$ is related to the equilibrium density of
  a single species by $n_{\alpha} = k_F^3/ 6 \pi^2$.
  Note that $\chi$ is independent of $a$, as expected from SU($N_c$) symmetry.
  Hence the matrix $\delta m_{\alpha \beta}$ and $\delta h_{\alpha \beta}^{\rm ext}$
  are simply proportional to each other.
  $M^*$ can be obtained by considering Galilean invariance, as in
  the two-component case \cite{LL}.
  We obtain
  \be
  M^*/M = 1 + N_c N(0) f_{1,s}/3
  \label{M*}
  \ee
   with
    $f_{1,s} \equiv 3 \langle (\hat k \cdot \hat k') f_s( \vec k, \vec k') \rangle $
    where $\hat k$ is the unit vector along $\vec k$ and the
    angular bracket denotes angular average.
     The speed of sound $u$ of the gas is related to $ \frac{d n}{ d \mu}  $ by
   $u^2 = \frac{n}{M} \left(\frac{ d n } { d \mu} \right)$, with $M$ the
   atomic (bare) mass.

\section{Dilute gas and $1/N_c$ expansions}

  So far we only discussed the general  formalism.  Now we specialize
  to the dilute limit and evaluate the quasiparticle interaction
  $f_{\alpha \gamma, \beta \delta} (\vec k, \vec k')$ to second order
  in the interaction $g$.  One way to proceed is to write down
  the total energy of the system up to $g^2$ and then take the
  derivative with respect to the occupation numbers (in the
  special case where all occupation numbers are diagonal), as done in
  \cite{LL} for the two-component system.  We immediately obtain
    \begin{widetext}
  \be
  f_{\alpha \alpha}(\vec k, \vec k')
   = - \left( \frac{ 4 \pi a} {M} \right)^2
   \left[ \frac{1}{V} \sum_{\vec k_1, \vec k_2, \gamma \ne \alpha}
  \frac{n_{\gamma} (\vec k_2) - n_{\gamma} (\vec k_1)}
  { \frac{k_2^2 - k_1^2}{2 M}} \delta_{\vec k_2 + \vec k', \vec k_1 + \vec k}
  \right]
  \label{fpp}
  \ee
  and
  \be
  f_{\alpha \beta \ne \alpha}(\vec k, \vec k')
   = \left( \frac{ 4 \pi a} {M} \right)
   - \left( \frac{ 4 \pi a} {M} \right)^2
   \left[ \frac{1}{V} \sum_{\vec k_1, \vec k_2}
  \frac{n_{\alpha} (\vec k_2) - n_{\beta} (\vec k_1)}
  { \frac{k_2^2 - k_1^2}{2 M}} \delta_{\vec k_2 + \vec k', \vec k_1 + \vec k}
  +
  \frac{1}{V} \sum_{\vec k_1, \vec k_2}
  \frac{n_{\alpha} (\vec k_1) + n_{\beta} (\vec k_2)}
  { \frac{2 k_F^2 - k_2^2 - k_1^2}{2 M}} \delta_{\vec k_1 + \vec k_2, \vec k + \vec k'}
  \right]
  \label{fpm}
  \ee
  \end{widetext}
  Here $V$ is the volume, $a$ the s-wave scattering length, and
  $n_{\gamma} (\vec k)$ is the equilibrium occupation number at $\vec k$ for
  $\gamma$ species (and thus equals unity if $k < k_F$ and zero otherwise).
  Compared with the spin $1/2$ system, where we shall denote the
  interaction between identical species as $f_{++}^{(1/2)} (\vec k, \vec k')$
  and distinguishable species as $f_{+-}^{(1/2)} (\vec k, \vec k')$ as in \cite{LL},
  we see that
  \be
  f_{\alpha \alpha} (\vec k, \vec k') = (N_c -1) f_{++}^{(1/2)} (\vec k, \vec k')
  \label{fppN}
  \ee
  and
    \be
  f_{\alpha \beta \ne \alpha} (\vec k, \vec k') =  f_{+-}^{(1/2)} (\vec k, \vec k')
  \label{fpmN}
  \ee
 and thus they can be directly obtained from the results for the spin $1/2$ system.

\begin{figure*}[tbp]

\begin{center}
\includegraphics[width=0.85\textwidth]{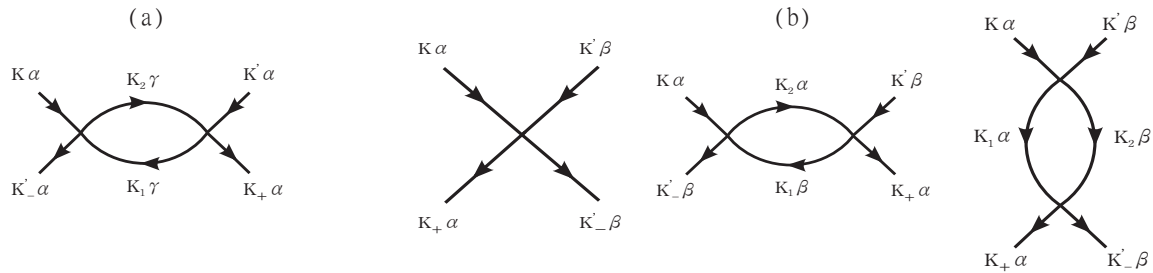}
\caption{{\bf Quasiparticle interactions in terms of Feynman diagrams.}
$f_{\alpha \alpha} (\vec k, \vec k')$ and
$f_{\alpha \beta \ne \alpha} (\vec k, \vec k')$
to second order
in $g$.
Notations are given in text.
}
\label{fig1}
\end{center}
\end{figure*}

 It is however more instructive to understand eq (\ref{fppN}) and (\ref{fpmN})
  in terms of Feymann diagrams.  $f_{\alpha \gamma, \beta \delta} (\vec k, \vec k')$
  can be expressed in terms of a special limit of the four point vertex function
  $\Gamma$ via \cite{LL,Nozieres}
  \be
  f_{\alpha \gamma, \beta \delta} (\vec k, \vec k') =
  Z^2 \ {\lim \atop \omega \to 0} \ {\lim \atop \vec q \to 0}
  \Gamma_{\alpha \gamma, \beta \delta} ( K+Q, K'-Q, K, K')
  \label{Gamma}
  \ee
  where we have used the short-hands $Q \equiv ( \vec q, \omega)$,
  $K \equiv (\vec k, \epsilon)$ etc with implicitly $\epsilon \to 0$,
  $k \to k_F$ etc.  Up to second order,
  $f_{\alpha \alpha} (\vec k, \vec k') = f_{\alpha \alpha, \alpha \alpha} (\vec k, \vec k')$
  is given solely by the diagram in Fig \ref{fig1}a, which gives
  the expression written in eq (\ref{fpp}).
  $f_{\alpha \beta \ne \alpha} (\vec k, \vec k') = f_{\alpha \beta, \alpha \beta} (\vec k, \vec k')$
  has a Hartree contribution in first order,
   and two second order contributions. They are depicted
    by the diagrams in Fig {\ref{fig1}}b.
   These diagrams correspond respectively
   to the three terms written down in eq (\ref{fpm}).  (The quasiparticle weight $Z$ can be
   taken simply as unity since the correction is at least second order in $a$).
   The $N_c -1$ factor enhancement of $f_{\alpha \alpha}$ as compared with the spin $1/2$ system
   is due to the increase in number of choices for the intermediate line labeled $\gamma$
   in Fig \ref{fig1}a.  It is also clear from Fig \ref{fig1}b that
   the effective interaction between distinguishable species is independent of $N_c$ to this order in $g$.
   We also note that Fig \ref{fig1}a has the from of an interaction arising from
   an induced density fluctuation in the $\gamma$ component, as is also clear
   from the appearance of a density response function (the term in the square bracket
   in eq (\ref{fpp})), though however it involves an exchange in the sense
   that the incoming $\vec k$ line has become an outgoing $\vec k'$ line, and vice
   versa.    (The corresponding diagram without the exchange vanishes
   under the limits in  eq (\ref{Gamma}))
    This exchange is responsible for the result that $f_{\alpha \alpha} (\vec k, \vec k') > 0$
   (the term in the square bracket in eq (\ref{fpp}) is negative definite), and thus an
   effective repulsive interaction between identical species.
   Similar induced interaction
   has been discussed in other contexts in cold atomic gases(e.g. \cite{Yu}).
   Note that in mean-field theory,
   $f_{\alpha \alpha} = 0$ since identical particles do not interact,
   whereas $f_{\alpha \beta}$
  is given entirely by the Hartree contribution $4 \pi a/M$.
   We shall see that corrections to mean-field results can be important
   in particular for large $N_c$.

  The expressions for $f_{++}^{(1/2)}$ and $f_{+-}^{(1/2)}$ can be found in
  textbooks.   To save space we would not reproduce these equations,
  as well as the results for $f_{\alpha \alpha}$ and $f_{\alpha \beta \ne \alpha}$ here,
  but relegate them to Supplemental Material \cite{Supp}.
  We directly give the results here for the effective mass
  \be
  M^*/M = 1 + (N_c -1) \frac{8}{15 \pi^2} [ 7 \ln 2 - 1 ] ( k_F a)^2 \ ,
  \label{Meff}
  \ee
  the inverse compressibility
  \be
  \frac{(d n / d \mu )^{-1}}{ (d n / d \mu)_{\rm free}^{-1}}
   = 1 + (N_c -1) \frac{2 k_F a}{\pi} [ 1 + \frac{ 2 k_F a} { 15 \pi} (22 - 4 \ln 2 ) ]
   \label{comp}
   \ee
   and the inverse generalized spin-susceptibility
   \be
   \frac{ \chi^{-1} }{ \chi_{\rm free}^{-1} }
    = 1 - \frac{2 k_F a}{\pi} - \frac{8 (k_F a)^2}{15 \pi^2} [ ( 11 - \frac{7 N_c}{2} )
      + 2 (N_c -1 ) \ln 2 ]
      \label{chi}
      \ee
   where the subscript ${\rm free}$ denotes the non-interacting gas.
   The compressibility can also be obtained from the energy per particle
   given in \cite{FW1,FW2}.

\begin{figure*}[tbp]
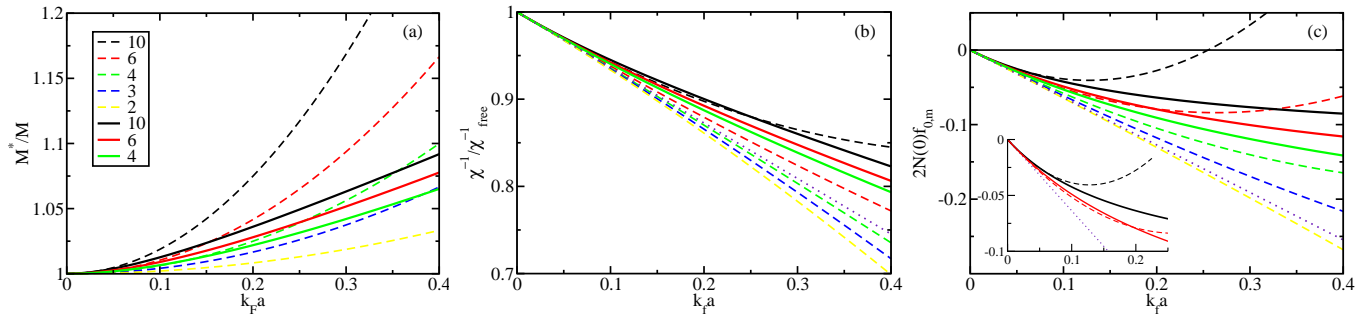

\begin{center}
\includegraphics[width=59mm]{fig2-mass.eps}
\includegraphics[width=59mm]{fig2-chi.eps}
\includegraphics[width=59mm]{fig2-f0m.eps}
\caption{ 
{(Color online)
(a) The effective mass $M^*/M$,
(b) inverse spin susceptibility,  as compared with its value for the free gas,
 and
(c) $2 N(0) f_{0,m}$, as functions of $k_F a$.}
Dashed lines: perturbation results to second order in $k_F a$; full lines: $1/N_c$ expansion.
Also shown in (b) and (c) are the mean-field results (dotted).
The inset in (c) show the details near small $k_F a$ for $N_c = 6$ and $10$.
}
\label{fig2}
\end{center}
\end{figure*}

   The results for $M^*/M$ and $\chi^{-1}/\chi^{-1}_{\rm free}$ are plotted as dashed lines in
   Fig \ref{fig2}.  The deviation of $M^*/M$ from unity is entirely due to
   non-mean-field effects,  and is larger for larger $N_c$.
  The modification of $dn/d \mu$ from its free gas value however is
  dominated by the mean-field correction (not plotted).
  The spin susceptibility would be discussed in more detail below.
   The compressibility and the spin susceptibility can be readily measured
   in experiments.  Both the compressibility and the magnetic susceptibility can be directly obtained from
   the density profile as a function of trap position
   if the local density approximation applies, and,  if the trap is harmonic,
   even the equation of state can be directly deduced (e.g. \cite{trap-t1,trap-t2,trap-ex1,trap-ex2,trap-ex3,trap-ex4,trap-ex5}).
   They can also be measured via
   number fluctuations of the gas. \cite{fluct-ex1,fluct-ex2}
   For example, for any two components
   $\alpha$ and $\beta$, the total relative number fluctuation within a subvolume $V$
   is related to $\chi$ via
   \be
   \langle (N_{\alpha} - N_{\beta} )^2 \rangle = k_B T V \chi
   \label{f-d}
   \ee
   where $k_B$ is the Boltzmann constant and $T$ the temperature.  Here
   the angular brackets denote averages over measurements.  More general expressions
   can be found in Supplemental Material \cite{Supp}.
  (In \cite{fluct-ex1,fluct-ex2}, seemingly only the fluctuation integrated along
    the line of sight were reported.  In principle, one can deduce also the local
    susceptibilities, in direct analogy to what has been done for the particle
    densities.  Details can also be found in Supplemental Material \cite{Supp}.)
   $M^*/M$ can in principle also be obtained from entropy via the equation of
   state \cite{CCLevin}

   Recently, there is a lot of interest on the ferromagnetic properties
   of the two component repulsive gas \cite{fluct-ex1,fluct-ex2,ferro-ex1,ferro-ex2,ferro-ex3,Duine,MC1,MC2,ferro-t1,ferro-t2,ferro-t3,ferro-t4,ferro-t5}.
   Hence,  we here would like to give here a more detailed discussion on
   the spin-susceptibility and contrast the
   $N_c = 2$ case with general $N_c$.  Fig \ref{fig2}c plots, in dashed lines,
   the term $2 N(0) f_{0,m}$ which
   appears in the denominator of $\chi$.  A negative (positive) sign of this
   term indicates an enhancement (suppression) of ferromagnetic tendency of the system.
    Within the mean-field approximation, it is simply
   $ - 2 k_F a / \pi$, and hence the susceptibility diverges
   at $k_F a = \pi/2$.  This divergence is independent of $N_c$, as also
   noted in \cite{CHU}.  However, the higher order terms in $k_F a$ modify this result.
   The second order contribution can be easily evaluated to be
   \bdm
   - \frac{2 k_F a}{\pi} \frac{ 4 k_F a}{3 \pi}
   \left[ ( 2 - \frac{N_c}{2}) - (N_c -1 ) \ln 2 \right]
   \edm
   For $N_c = 2$, this term is negative, thus the
   ferromagnetic tendency is enhanced. \cite{Duine}  (This qualitative trend is
   supported by numerical evaluations \cite{MC1,MC2}). However, for $N_c \ge 3$,
    the sign of this term is already reversed.
   This is a consequence of the fact that the effective
   repulsive interaction between identical species is proportional to $N_c -1$
   and hence enhanced.
   Using the above formulas,  we find that, for $N_c \ge 3$,
   $2 N(0) f_{0,m} > -1$ for arbitrary large $k_F a$, and hence
      there is no ferromagnetic instability under our approximations
   (though a first order phase transition \cite{CHU,Duine,Heiselberg11} to
   a ferromagnetic state is still possible).
   We note that, as can be seen from Fig 2, the correction to mean-field can already be
  substantial even for modest $k_F a$ for large $N_c$.

     Our expansion in  $k_F a$  gives
   a large and positive $2 N(0) f_{0,m}$  for large $k_F a$ and $N_c \ge 3$.  However,
   this result is likely to be an artifact of the $k_F a$ expansion.
   In fact, one expects that the range of $k_F a$ values
   for our expressions above to be reliable should be restricted to $k_F a < 1/ N_c$,
   and hence become very small for $N_c$ such as $6$ or $10$.
   Hence we consider also the alternative where one
   regards $ (N_c k_F a)$ as a new (fixed) parameter, and to perform
   an expansion in $1/N_c$.   The details are given in Supplemental Material \cite{Supp}.
   Both $f_{\alpha \alpha}$ and $f_{\alpha \beta}$
   are formally of order $1/N_c$.  They are both finite for large $N_c k_F a > 0$
   as a result of screening.
   The corresponding results for $2 N(0) f_{0,m}$ are also plotted as full lines in Fig \ref{fig2}a.
  We see that in general this factor is negative,
  saturates at large $k_F a$, and to a smaller value for larger $N_c$.
  The  results for $\chi^{-1}$, normalized to their free gas values,
  are shown as full-lines in Fig \ref{fig2}b.  It decreases fast with $k_F a$, mainly
  due to the increase in $M^*/M$.

  The behaviors of $\chi$ and $f_{0,m}$ are often considered as indicators of
  whether there is a ferromagnetic tendency of the system.
  If we examine these quantities alone, our results above would indicate
  that, for larger $N_c$, the system is further away from ferromagnetism.
  Interestingly however, an examination of the energy of the system
  indicates that the situation is more complex.  Including
  the interaction energy up to first or second order in $a$, it can be
  shown \cite{Heiselberg11,PC} that the unpolarized state actually becomes unstable
  at a smaller $k_F a$ for larger $N_c$.  This may indicate that,
  for larger $N_c$, the ferromagnetic transition actually becomes more
  strongly first order.  It would be of great interest to examine
  whether this remains true when higher order interaction terms
  are included.


\section{Conclusions}

   In summary, we have considered the Fermi liquid theory of an SU(N)
   dilute gas.  Fermi liquid parameters are evaluated beyond the mean field
   approximation.  Though the corrections to mean-field are formally higher
   order in the gaseous parameter $k_F a$, they are nonetheless enhanced
   by the number of components $N_c$.  Currently, $k_F a$ in the $^{173}$Yb experiments is
   $\approx 0.13$ with $N_c$ as large as $6$.  While $k_F a$ for
   the $^{87}$Sr experiments are currently somewhat smaller, $N_c$ can be as large as $10$.
   In both cases, these beyond mean-field effects should be easily measurable,
   and they would be even more important if experiments can be carried out
   at higher densities.

\section{Acknowledgements}

  This research is supported by the National Science Council of Taiwan under
  grant number NSC 101-2112-M-001 -021 -MY3.  Part of this work was performed
  while SKY was in NORDITA and the Aspen Center for Physics.  SKY would
  like to acknowledge their hospitality, as well as the support
  by the National Science Foundation under Grant No. PHYS-1066293 and
  the Simon Foundation during his visit in Aspen.
  We would also like to thank Miguel Cazalilla for helpful communications.




\end{document}




\title{Theory of SU(N) Fermi liquid
\\ Supplemental Material\\
}


\author{S.-K. Yip}
\affiliation{Institute of Physics, Academia Sinica, Taipei 115, Taiwan}
\affiliation{Institute of Atomic and Molecular Sciences, Academia Sinica, Taipei 106, Taiwan}
\author{Bor-Luen Huang}
\affiliation{Institute of Physics, Academia Sinica, Taipei 115, Taiwan}
\author{Jung-Shen Kao}
\affiliation{Institute of Physics, Academia Sinica, Taipei 115, Taiwan}
\affiliation{Department of Physics, National Taiwan University, Taipei 106, Taiwan}


\date{\today}

\maketitle

\subsection{derivation of formulas for compressibility and susceptibility}

The total change in quasiparticle energy is
$\delta \epsilon_{\alpha \beta}^{\rm tot} =
\epsilon_{\alpha \beta}^{\rm ext} + \delta \epsilon_{\alpha \beta}$ with
the last term being the contribution from interaction.
Expanding these quantities as in eq (6) and (7) in text and using (8) and (9),
we get
\begin{widetext}
\be
\frac{\delta n}{N_c} \delta_{\alpha \beta} +
\frac{1}{2} \sum_a m^a T^a_{\alpha \beta}
= - N(0) \left\{ \left[ \epsilon_s^{\rm ext} + f_{0,s} \delta n \right] \delta_{\alpha \beta}
  - \sum_a \left[ \left( h^{a, {\rm ext}} - f_{0,m} m^a \right) T^a_{\alpha \beta} \right] \right\}
  \ee
  \end{widetext}
from which we can deduce (14) and (15).

\subsection{collection of formulas in $(k_F a)$ expansion:}

For spin $1/2$, we have, from \cite{LL},
\be
f_{++}^{(1/2)} (\vec k, \vec k') = \frac{4 \pi a} { M} \frac{k_F a}{\pi} \Phi_1(s)
\ee
\be
f_{+-}^{(1/2)} (\vec k, \vec k') = \frac{4 \pi a} {M} \left[ 1 +
\frac{k_F a}{\pi}  ( \Phi_1 (s) + \Phi_2 (s) ) \right]
\ee
where $s = \sin (\chi/2)$ with $\chi$ the angle between $\vec k$ and $\vec k'$,
and
\be
\Phi_1 (s) = 1 + \frac{ 1 - s^2}{2 s} \ln \frac{ 1 + s} { 1 -s }
\ee
\be
\Phi_2 (s) = 2 \left[ 1 - \frac{s}{2} \ln \frac{ 1 + s} { 1 -s } \right]
\ee
Correspondingly, we have (see also \cite{Recati11})
\be
f_{0,++}^{(1/2)} = \frac{ 2 \pi a} {M}
\left[ \frac {4 k_F a}{3 \pi}\right] \left[  1 + 2 \ln 2  \right]
\ee
\be
f_{0,+-}^{(1/2)} = \frac{ 2 \pi a} {M}
\left[ 2 + 3 \frac {4 k_F a}{3 \pi}\right]
\ee

\be
f_{1,++}^{(1/2)} = \frac{ 2 \pi a} {M}
\left[ \frac {4 k_F a}{5 \pi}\right] \left[  6 \ln 2   - 3   \right]
\ee
\be
f_{1,+-}^{(1/2)} = \frac{ 2 \pi a} {M}
\left[ \frac {4 k_F a}{5 \pi}\right] \left[ 8 \ln 2 + 1 \right]
\ee

For the $N_c$ component system,
we get
\be
f_{0,s} = \left( 1 - \frac{1}{N_c} \right)
 \frac{ 2 \pi a} {M}
\left[ 2 +  \frac {4 k_F a}{3 \pi} \left( 4 + 2 \ln 2  \right) \right]
\ee
\be
f_{0,m} = - \frac{ 2 \pi a} {M}
\left[ 1 +  \frac {4 k_F a}{3 \pi} \left(  ( 2 - \frac{N_c}{2} ) -
(N_c -1 ) \ln 2  \right) \right]
\ee
\be
f_{1,s} = ( 1 - \frac{1}{N_c}) \frac{ 2 \pi a} {M} \frac{8 k_F a}{5 \pi^2} [ 7 \ln 2 - 1 ]
\ee
\be
f_{1,m} = \frac{\pi a }{M} \frac{ 4 k_F a}{ 5 \pi} \left[
  ( 6 N_c - 14 ) \ln 2 - ( 3 N_c - 2) \right]
  \ee
From these results, we obtain the compressibility and generalized spin-susceptibility in
text.

\subsection{fluctuations:}

Generally, we have
\be
\langle N_{\alpha} N_{\beta} \rangle - \langle N_{\alpha}  \rangle \langle  N_{\beta} \rangle
 = - k_B T V \left( \frac{ \partial n_{\alpha} } { \partial \epsilon_{\beta}^{\rm ext} } \right)
 \ee
 The partial derivative on the right-hand-side of this equation can be evaluated
 as follows.  Consider general external potential
 $\delta \epsilon_{\alpha}^{\rm ext}$  diagonal in the species index $\alpha$.
  We separate out a part which is proportional
 to the average over $\alpha$. That is, we define
 $\epsilon_s^{\rm ext} = \frac{1}{N_c} \sum_{\alpha} \epsilon_{\alpha}^{\rm ext}$
 and write $\epsilon_{\alpha}^{\rm ext} = \epsilon_s^{\rm ext}
 - h_{\alpha}^{\rm ext}$.  Note that $\sum_{\alpha} h_{\alpha}^{\rm ext} = 0$ by definition.
 Similarly, we have
 $n_{\alpha} = \frac{n}{N_c} + \frac{1}{2} m_{\alpha}$, where $n$ is the total density
 $\sum_{\alpha} n_{\alpha}$.  Again $\sum_{\alpha} m_{\alpha} = 0$.
 We know
 $dn = - \frac{ d n} { d \mu} d \epsilon_s^{\rm ext} $ and
 $m_{\alpha} =  \chi h_{\alpha}^{\rm ext}$.
 Hence
 \be
 \delta n_{\alpha} = - \frac{1}{N_c} \left( \frac{ d n} { d \mu} \right)  \epsilon_s^{\rm ext}
  + \frac{1}{2} \chi h_{\alpha}^{\rm ext}
  \ee
  Rewriting $\epsilon_s^{\rm ext}$ and $h_{\alpha}^{\rm ext}$ in terms of
  $\epsilon_{\alpha}^{\rm ext}$, we get
  \be
  \frac{ \partial n_{\alpha} }{ \partial \epsilon_{\alpha}^{\rm ext}}
   = - \frac{1}{N_c^2} \left( \frac{ d n} { d \mu} \right) -
   \frac{1}{2} \chi \left( 1 - \frac{1}{N_c} \right)
   \ee
   and, for $\beta \ne \alpha$,
    \be
  \frac{ \partial n_{\alpha} }{ \partial \epsilon_{\beta}^{\rm ext}}
   = - \frac{1}{N_c^2} \left( \frac{ d n} { d \mu} \right) +
   \frac{1}{2 N_c} \chi
   \ee
   with $\frac{ d n} { d \mu}$ and $\chi$ already evaluated in text.
   From these formulas we can deduce that
   \be
   \langle N^2 \rangle - \langle N \rangle^2
    = k_B T V \left( \frac{ d n} { d \mu} \right)
    \ee
    as usual, and, if $\beta \ne \alpha$,
    \be
    \langle ( N_{\alpha} - N_{\beta} )^2 \rangle = k_B T V \chi
    \ee
    Generally,  we obtain, if $c_{\gamma}$ are a set of real numbers,
    \begin{widetext}
    \be
    \langle (\sum_{\gamma} c_{\gamma} N_{\gamma})^2 \rangle -
     (\langle \sum_{\gamma} c_{\gamma} N_{\gamma} \rangle)^2
     = k_B T V \left\{ (\bar{c})^2  \left( \frac{ d n} { d \mu} \right)
     + \frac{1}{2} \left[ \sum_{\gamma} (c'_{\gamma})^2 \right] \chi \right\}
     \ee
     \end{widetext}
     where $\bar c \equiv \frac{1}{N_c} \sum_{\gamma} c_{\gamma}$
     is the average of $c_{\gamma}$, and $c'_{\gamma} \equiv c_{\gamma} - \bar c$.
     Note that $\sum_{\gamma}  c'_{\gamma} = 0$.

   The above is for the fluctuation in a volume where the parameters (such as density)
   can be taken as constant.  In typical cold atom experiments, however, the density is a function of
   position.  Let us consider a cylindrically symmetric trap, with potential of
   the form $V (\vec r) = \frac{M}{2} ( \omega^2 \rho^2 + \omega_z^2 z^2)$, with
   $\omega$ and $\omega_z$ the trap frequencies.  $\rho^2 = x^2 + y^2$.
   Suppose that, in a particular experiment, one measures the number fluctuation along
   a line of sight $x$ and cross-section $(\Delta y)( \Delta z)$, then we have
   \be
   \langle ( N_{\alpha} - N_{\beta} )^2 \rangle_{\rm col} =
   k_B T (\Delta y)( \Delta z) \chi_{\rm col} (y,z)
   \ee
   where
   \be
   \chi_{\rm col} (y, z)  = \int dx \chi (x, y, z)
   \ee
   is a column integrated susceptibility.  Due to
   cylindrical symmetry, $\chi (x, y,z)$ depends on $x$ and $y$ only through $\rho$, hence we have
   \be
   \chi_{\rm col} (y, z) = \int_{y}^{\infty} d \rho \frac{2 \rho}{\sqrt{ \rho^2 - y^2}} \chi( \rho, z)
   \ee
    Using the same
   argument as for the column integrated densities,  the local
   density can be obtained via an inverse Abel transform \cite{Abel}
   \be
   \chi( \rho, z) = - \frac{1}{\pi} \int_{\rho}^{\infty} dy \frac{ d \chi_{\rm col} (y,z)/ d y}
                {\sqrt{ y^2 - \rho^2}}
   \ee
   Additional relations can be found if the trap is harmonic.
   If we integrate also along $y$, we obtain
   the "axial" susceptibility $\chi_{\rm axi} (z) = \int dy \chi_{\rm col} (y, z)
    = 2 \pi \int d \rho \rho \chi (\rho, z)$, related to the total fluctuation in number difference
    within a disc of thickness $\Delta z$:
    \be
    \langle (N_{\alpha} - N_{\beta})^2 \rangle_{\rm disc} = k_B T (\Delta z) \chi_{\rm axi} (z)
    \ee
     If local density approximation is
    assumed, we have also \cite{trap-t}
    \be
    \chi (0, z) = - \frac{1}{2 \pi z} \frac{ \omega^2} {\omega_z^2}
    \frac{ d \chi_{\rm axi} (z)} { dz}
    \ee
    hence the value of the susceptibility $\chi$ of the system at the local density $n (0, z)$.

\subsection{ $1/N_c$ expansion:}

\begin{figure}[tbp]
\includegraphics[width=70mm]{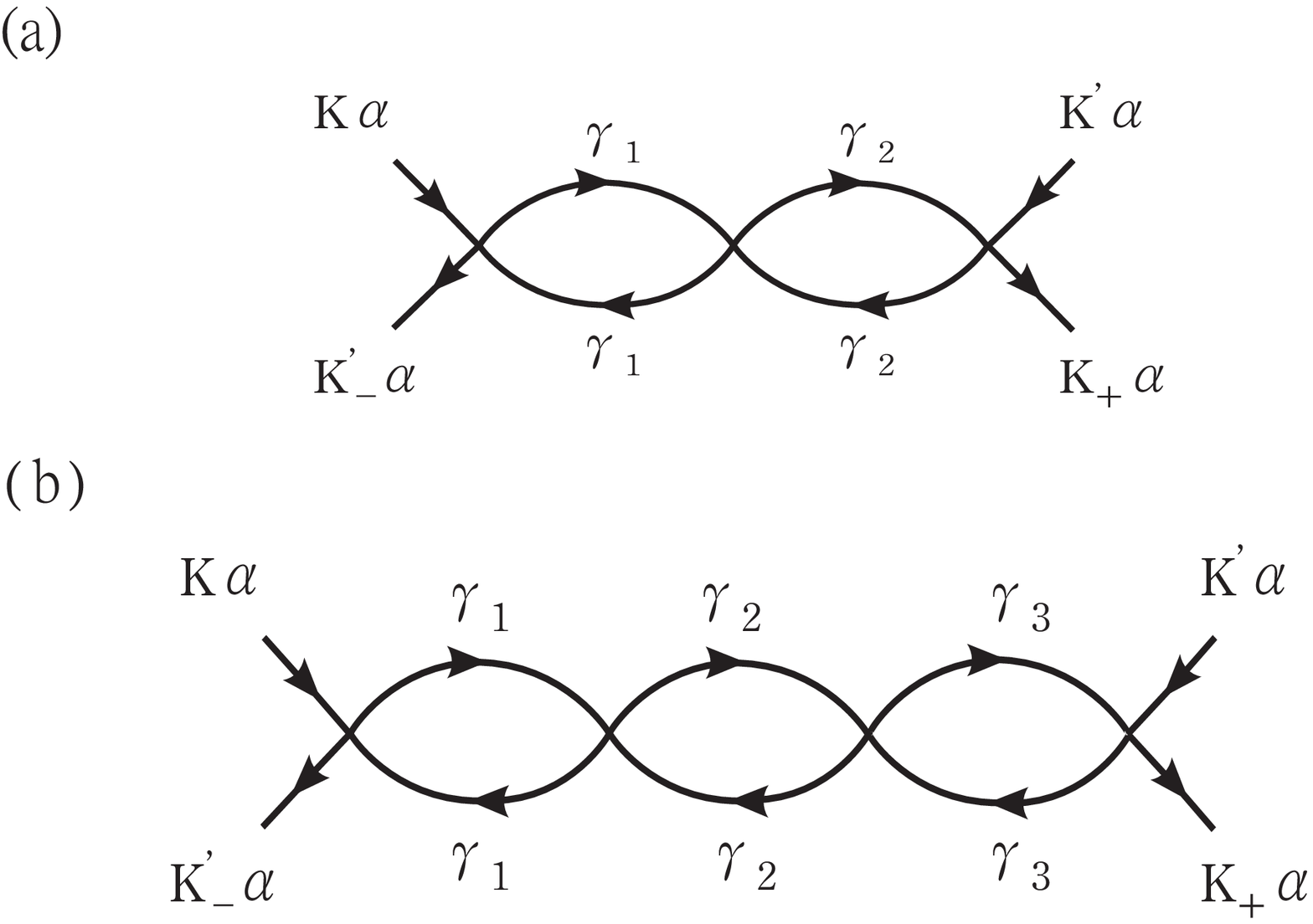}
\caption{Dominant third and fourth order contributions to $f_{\alpha \alpha} (\vec k, \vec k')$
in $(4 \pi k_F a /M)$ for large $N_c$.
}
\label{sup-fig1}
\end{figure}

  In principle we can perform an $1/N_c$ expansion using a more formal
  field-theoretical method, but in below we choose to provide our calculation
 in a manner more align with our presentation in text.
 We have seen that, to order $(k_F a)^2$, Fig 1a  in text is the only term in
   the quasiparticle interactions that increases with $N_c$.  It only
   contributes to $f_{\alpha \alpha}$ and is  proportional to $(N_c -1)$.
    Let us consider higher order terms, and for the moment, exclude those diagrams
    which correspond to insertion of Hartree self-energy corrections to the lower order diagrams.
   To order $(k_F a)^3$ and $(k_F a)^4$, it can be easily seen that
   the terms that increases fastest with $N_c$ are given in Fig \ref{sup-fig1}a and b
   respectively, and they are all for $f_{\alpha \alpha}$.  The diagram in Fig \ref{sup-fig1}a
   is proportional to $(N_c -1 ) (N_c -2)$ and hence $N_c^2$ for large $N_c$.
   It can be checked that there are no diagrams that are $\propto N_c^2$ to this order
   in $(k_F a)$.  Similarly the diagram in Fig \ref{sup-fig1}b is proportional to
   $ (N_c -1) ( (N_c -1) + (N_c -2 )^2) = (N_c -1) ( N_c^2 - 3 N_c + 3)  $ \cite{counting} and increases as $N_c^3$ for
   large $N_c$, and all other diagrams to  order $g^4$ can be seen to increase with
   $N_c$ slower than $N_c^3$.  Similar statements can be made for higher order terms in $g$.
   Summing over these diagrams with the highest powers in $N_c$ give the series
   \begin{widetext}
   \be
   f_{\alpha \alpha} (\vec k, \vec k')
   = (N_c - 1) \left( \frac{ 4 \pi a } {M} \right)^2 L (\vec k - \vec k')
   \left\{ 1 - (N_c-2) \left( \frac{ 4 \pi a } {M} \right) L (\vec k - \vec k')
     + (N_c^2 - 3N_c + 3) \left[ \left( \frac{ 4 \pi a } {M} \right) L (\vec k - \vec k') \right]^2
     + \ldots \right \}
   \ee
   \end{widetext}
   where $L (\vec k - \vec k') = - \frac{1}{V} \sum_{\vec k_1} \frac{n(\vec k_1 + \vec k - \vec k') - n(\vec k_1)}
   {  [(\vec k_1 + \vec k - \vec k')^2 - \vec k_1^2] / 2 M }$ is the Lindhard function.
   This series sums to,  if we ignore the differences between $(N_c -1)$ and $(N_c -2)$ etc
   \bdm
  \frac{ N_c \left( \frac{ 4 \pi a } {M} \right)^2 L (\vec k - \vec k')}
  { 1 + N_c \left( \frac{ 4 \pi a } {M} \right) L (\vec k - \vec k')}
  \edm
  The denominator just represents basically a screening
  which makes the result finite rather than diverging in the large $N_c k_F a$ limit.
    Inserting back the Hartree corrections to the bare one particle Green's function lines
   does not modify $L (\vec k - \vec k')$, and hence the above result.
  $L( \vec k - \vec k') $ can be evaluated to be
  $\frac{ M k_F }{ 4 \pi^2} \Phi_1 (s)$.
  Hence we get
  \be
  f_{\alpha \alpha} (\vec k, \vec k') = \frac{1}{N_c}
  \frac{ 4 \pi N_c a} {M}
  \frac{ \frac{ N_c k_F a}{\pi} \Phi_1 (s) }
  { 1 + \frac{ N_c k_F a}{\pi} \Phi_1 (s) }
  \ee
  This result is valid to order $1/N_c$.  For the effective interaction
  between distinguishable particles, the Hartree contribution is
  already of order $1/N_c$, thus
  \be
  f_{\alpha \beta \ne \alpha} (\vec k, \vec k') = \frac{1}{N_c}
  \frac{ 4 \pi N_c a} {M}
  \ee
  with corrections of order $1/N_c^2$.

Using $2 f_{0,m} =   f_{0, \alpha \alpha} - f_{0, \alpha \beta \ne \alpha} $ we obtain,
for the denominator for the susceptibility,
  \be
  1 + 2 N(0) f_{0,m}
  = 1 - \frac{1}{N_c} \frac{ 2 N_c k_F a} {\pi}
  \langle \frac{1} { 1 + \frac{N_c k_F a} {\pi} \Phi_1 (s) } \rangle
  \ee
  which is correct to order $1/N_c$. This result is plotted in Fig 2c in text.
  To obtain $M^*/M$ to the same order however, a complication arises.
  Since $M^*/M = 1 + N(0) N_c f_{1,s}/3 = $
   $ 1 + N(0) [ f_{1, \alpha \alpha} + (N_c -1) f_{1, \alpha \beta \ne \alpha} ]/ 3$,
   we need $f_{1, \alpha \beta \ne \alpha}$ correct up to order $1/N_c^2$.
   These diagrams are complicated to calculate and we would not perform them here.
   We know however that $f_{\alpha \beta}$ up to order $a^2$ can be written as
   \be
f_{\alpha \beta \ne \alpha} (\vec k, \vec k') = \frac{1}{N_c} \frac{ N_c 4 \pi a} {M} \left[ 1 +
\frac{1}{N_c} \frac{N_c k_F a}{\pi}  ( \Phi_1 (s) + \Phi_2 (s) ) \right]
\ee
with the second term of order $1/N_c^2$.
From argument similar to that given for $f_{\alpha \alpha}$ above, we expect that diagrams
higher order in $a$ but still of order $1/N_c^2$ leads roughly to a screening of
the second term.   We shall make the {\it ansatz}
\be
f_{\alpha \beta \ne \alpha} (\vec k, \vec k') = \frac{1}{N_c} \frac{ N_c 4 \pi a} {M} \left[ 1 +
\frac{1}{N_c}
\frac{N_c k_F a}{\pi} \frac{ ( \Phi_1 (s) + \Phi_2 (s) )} {  1 + \frac{N_c k_F a} {\pi} \Phi_1 (s) } \right]
\ee
Writing this result as
$ \frac{1}{N_c} \frac{ N_c 4 \pi a} {M} \left[ 1 + \frac{1}{N_c} Y(s) \right]$ and defining
 $X(s) \equiv   \frac{ N_c k_F a}{\pi} \Phi_1 (s) /
  [ 1 + \frac{ N_c k_F a}{\pi} \Phi_1 (s) ]$, we finally obtain
  $M^*/M = 1 + \frac{1}{N_c} \frac{ 2 N_c k_F a}{ \pi}
  \langle ( 1 - 2 s^2 ) (X(s) + Y(s)) \rangle$
  and hence $\chi^{-1}/\chi^{-1}_{\rm free}$ in Fig 2a and 2b in text.

